# Scalable Systems and Software Architectures for High-Performance Computing on cloud platforms

Risshab Srinivas Ramesh
*Computer Science Engineering*
*Ramaiah Institute of Technology*
*Bangalore, India*
risshabsrinivas@gmail.com

*Abstract*—High-performance computing (HPC) is essential for tackling complex computational problems across various domains. As the scale and complexity of HPC applications continue to grow, the need for scalable systems and software architectures becomes paramount. This paper provides a comprehensive overview of architecture for HPC on premise focusing on both hardware and software aspects and details the associated challenges in building the HPC cluster on premise. It explores design principles, challenges, and emerging trends in building scalable HPC systems and software, addressing issues such as parallelism, memory hierarchy, communication overhead, and fault tolerance on various cloud platforms. By synthesizing research findings and technological advancements, this paper aims to provide insights into scalable solutions for meeting the evolving demands of HPC applications on cloud.

## I. Introduction

High Performance Computing (HPC) architectures consist of several key building blocks that work together to create a powerful and scalable computing environment capable of tackling complex computational tasks. These building blocks can be categorized into hardware components, software components. Let's explore each of these building blocks and how they are used to create an HPC architecture:

## II. On Premise HPC Architecture

High-Performance Computing (HPC) is a method of processing large amounts of data and performing complex calculations at high speeds. On-premises HPC architecture involves setting up and managing HPC systems within an organization's own data center.

### A. Hardware Components

- Compute Nodes: A cluster of computers, each with multiple processors, dedicated circuits, and local memory, performing computations. Few examples include Intel Xeon Scalable processors, AMD EPYC processors, or ARM-based processors like Ampere Altra. These processors provide high core counts and advanced features like AVX-512 instructions for accelerating HPC workloads.

- GPU-Accelerated Systems: Graphics Processing Units (GPUs) provide parallel processing capabilities to support tasks that can be parallelized on both CPU cores and GPUs. Some of the GPU's include NVIDIA A100 or A40 GPUs, AMD Instinct MI100 or MI200 GPUs provide massive parallel processing capabilities to speed up AI, machine learning, and scientific computing tasks.

- High-Performance Storage: Parallel file systems or high-speed storage controllers enhance data access and transfer speeds. Some of the HPS include Lustre parallel file systems, IBM Spectrum Scale (GPFS), or BeeGFS. These file systems leverage multiple storage servers and high-speed networks to deliver extremely high bandwidth and IOPS.

- InfiniBand Switch: Connects and facilitates communication between all nodes in the cluster. Mellanox/NVIDIA InfiniBand switches like the Quantum 8700 series provide high-bandwidth, low-latency interconnects between compute nodes.

- Do Facilities and Power: Physical space required to accommodate HPC equipment, including power supply and cooling infrastructure.

- Field-Programmable Gate Arrays (FPGAs): Customizable hardware acceleration for environments requiring highly efficient and low-latency processing.

- Remote Visualization Nodes: Offload visualization tasks from main compute nodes to maintain computational efficiency.

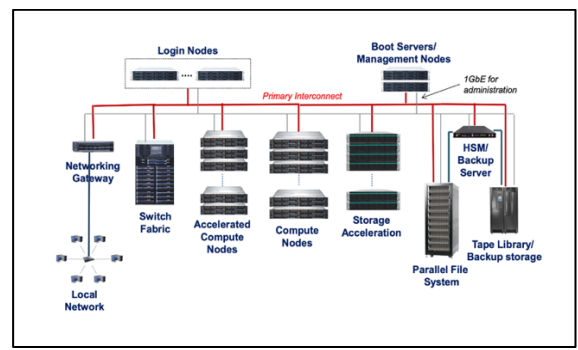

Fig. 1. Components of a cluster

### B. Software Components

- HPC Schedulers: Specialty software that orchestrates shared computing resources, ensuring efficient node utilization. Examples of schedulers include SLURM, and Univa Grid Engine. These schedulers efficiently manage shared compute resources and workloads.

- In Parallel Computing Software: Manages parallel tasks, ensuring efficient coordination and synchronization. Examples include OpenMP, MPI, CUDA, and OpenACC. These libraries and frameworks enable developers to parallelize their code to run efficiently on HPC systems.

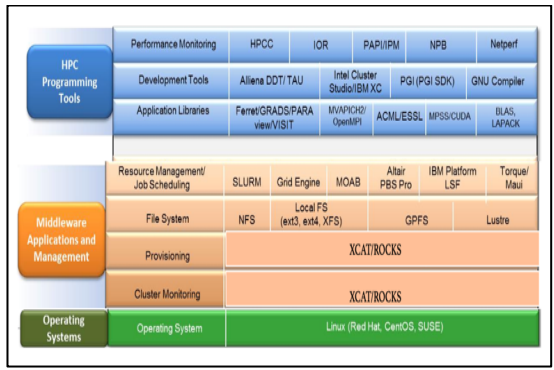

Fig. 2. HPC Software Stack

- Data Management Software: Optimizes system resource management, handles data storage, retrieval, organization and movement with the HPC environment. Examples are IBM Spectrum Scale, DataWarp from Cray and BeeGFS. These software-defined storage solutions optimize data placement, tiering, and movement.

### III. Key Challenges Of Implemeting On-Premises HPC

Setting up and maintaining an on-premises HPC system requires significant upfront and ongoing costs for

#### A. High Costs

- Hardware: Purchasing powerful servers, storage systems, networking equipment, and other necessary components can be extremely expensive.
- Skilled Technicians: Hiring and retaining qualified personnel to manage and maintain the HPC cluster is costly, as these individuals often have specialized skills and expertise.
- Facilities: Providing adequate space, power, cooling, and other infrastructure to support the HPC cluster can add significant costs to the overall operation.

#### B. Infrastructure Maintenance

- Node Failure: As HPC equipment ages, the likelihood of node failures increases, requiring regular maintenance and upgrades to ensure system reliability and performance.
- Upgrades: Keeping the HPC cluster up to date with the latest hardware and software technologies requires ongoing upgrades, which can be time-consuming and costly.
- Downtime: Performing maintenance and upgrades can lead to system downtime, which can impact productivity and delay research or business-critical tasks.

#### C. Long Purchasing Cycles

- High Demand: Due to the specialized nature of HPC equipment and the large number of organizations requiring these systems, there is often high demand, leading to longer purchasing cycles.
- Customization: Many HPC systems are customized to meet specific requirements, which can further extend the purchasing process.
- Vendor Availability: Depending on the vendor and the specific components needed, the availability of HPC equipment can vary, leading to delays in acquiring the necessary hardware.

#### D. Scalability

- Architectural Design: Creating a scalable HPC architecture that allows for seamless addition or removal of computational resources requires careful planning and design.
- Heterogeneous Components: Integrating different types of hardware and software components, such as CPUs, GPUs, and specialized accelerators, can make scalability more challenging.
- Software Compatibility: Ensuring that the software stack, including operating systems, libraries, and applications, is compatible with the added or removed resources is crucial for maintaining performance and stability.

#### E. Data Management

- Large Data Volumes: HPC workloads often generate and process massive amounts of data, which requires sophisticated networking and storage infrastructure to handle the data efficiently.
- Data Transfer: Moving large datasets between storage systems and compute nodes can be time-consuming and can impact overall system performance.
- Data Backup and Recovery: Implementing robust data backup and recovery strategies is essential to protect against data loss and ensure business continuity.

#### F. Power Consumption and Cooling

- High Heat Production: HPC systems generate significant amounts of heat due to the high-performance components and dense server configurations, necessitating specialized cooling technologies.
- Power Consumption: Running an HPC cluster requires a substantial amount of electricity, which can be expensive and difficult to maintain, especially in regions with high energy costs or limited power infrastructure.
- Cooling Costs: Providing adequate cooling for the HPC cluster can add significant costs to the overall operation, both in terms of equipment and energy consumption.

#### G. Software Challenges

- Version Management: Keeping track of the various software component versions, including operating systems, libraries, and applications, and ensuring compatibility between them can be challenging.
- Updates and Patches: Applying updates and patches to the software stack can be risky, as they may

introduce compatibility issues or impact system stability and performance.
- Customization: Many HPC applications require customization or the use of specialized software, which can add complexity to the software management process.

## IV. CREATING THE HPC ARCHITECTURE

To create an HPC architecture, these building blocks are assembled and configured to meet the specific requirements of the target application workload. Compute nodes are interconnected using high-speed interconnects and organized into a cluster. Storage systems are provisioned to store input data, intermediate results, and output data generated by computational tasks. The operating system, middleware, and development tools are installed and configured to support parallel computing, job scheduling, and application development. Networking infrastructure is configured based on the chosen network topology, with switches, routers, and network management tools deployed to ensure efficient communication between compute nodes. Finally, HPC applications are developed or ported to the architecture using parallel programming models and libraries, and performance optimizations are applied using development tools and profiling techniques to maximize computational efficiency.

Building scalable systems and software architectures for HPC presents numerous challenges and considerations. Among these challenges, load balancing, resilience, and energy efficiency play crucial roles in determining the overall performance and sustainability of HPC architectures.

### A. Load Balancing

Load balancing is essential for distributing computational tasks evenly across resources in a parallel computing environment. Non-uniform workload distribution can lead to resource underutilization and performance bottlenecks, hindering scalability. In HPC architectures, load balancing becomes particularly challenging due to the dynamic nature of workloads and the heterogeneity of computing resources.

Cloud platforms like AWS, Azure, and Google Cloud (GCP) offer various load balancing services and features to distribute computational tasks evenly across resources for HPC workloads:

AWS Elastic Load Balancing (ELB) automatically distributes incoming application traffic across multiple EC2 instances. ELB supports multiple load balancing algorithms, including round-robin, least outstanding requests, and least executed requests, to optimize load distribution. AWS Auto Scaling can be used in conjunction with ELB to automatically scale compute resources based on demand, ensuring efficient load balancing as workloads change.

Azure Load Balancer distributes inbound traffic among healthy backend instances, providing high availability and outbound connectivity for HPC applications. Azure Load Balancer supports various load balancing algorithms, such as hash-based distribution, five-tuple (source IP, source port, destination IP, destination port, protocol type), and source IP affinity. Azure Autoscale can automatically scale compute resources based on performance metrics or a schedule, enabling dynamic load balancing for HPC workloads.

GCP Load Balancing offers global and regional load balancing options for distributing traffic across multiple instances and regions. GCP Load Balancing supports various load balancing algorithms, including round-robin, least request, and random, to optimize load distribution. GCP Autoscaler can automatically scale compute resources based on load, ensuring efficient load balancing as workloads change.

These load balancing services and features help distribute computational tasks evenly across resources in HPC clusters on AWS, Azure, and GCP. By leveraging dynamic load balancing algorithms, auto-scaling capabilities, and support for heterogeneous computing resources, these cloud platforms can mitigate resource under-utilization and performance bottlenecks, enabling scalable and efficient execution of HPC workloads.

### B. Resilience

Resilience refers to the ability of an HPC system to maintain functionality and performance in the presence of hardware or software failures. As HPC systems scale to thousands or even millions of compute nodes, the likelihood of failures increases, posing significant challenges to system reliability and availability. Resilience mechanisms are crucial for minimizing the impact of failures on application execution and ensuring uninterrupted operation of HPC workloads. Cloud platforms like AWS, Azure, and Google Cloud (GCP) offer various resilience features to ensure high availability and fault tolerance for HPC workloads:

AWS Auto Scaling dynamically adjusts capacity and automatically launches or terminates EC2 instances based on user-defined policies, ensuring resilience and high availability of HPC applications. AWS Elastic Block Store (EBS) provides durable and scalable storage volumes with built-in redundancy and data replication, protecting against data loss and ensuring data availability for HPC workloads. AWS Batch enables running batch computing workloads on the AWS Cloud, automatically scaling compute resources based on the volume and resource requirements of the submitted jobs, enhancing resilience.

Azure Autoscale automatically scales out or in compute resources based on performance metrics or a schedule, ensuring resilience and efficient resource utilization for HPC applications. Azure Blob Storage and Azure Files offer scalable and redundant storage options with built-in data replication, protecting against data loss and ensuring data availability for HPC workloads. Azure Site Recovery provides disaster recovery capabilities, allowing organizations to replicate and failover HPC workloads to a secondary Azure region in case of a regional outage or disaster.

GCP Autoscaler automatically scales compute resources based on load, ensuring resilience and efficient resource utilization for HPC applications. Google Cloud Storage provides highly durable and available object storage with built-in redundancy and data replication, protecting against data loss for HPC workloads. GCP Dataflow offers a managed service for executing batch and streaming data processing pipelines, automatically scaling resources and handling failures, enhancing resilience for HPC applications.

These cloud platforms provide resilience features like automatic scaling, redundant storage, and managed services to ensure high availability and fault tolerance for HPC workloads. By leveraging these capabilities, organizations can minimize the impact of failures and ensure uninterrupted operation of their HPC applications, even in the face of hardware or software failures.

*C. Energy Efficiency*

Energy efficiency is becoming increasingly important in HPC architectures due to rising power consumption and cooling costs. Large-scale HPC systems consume substantial amounts of energy, leading to environmental concerns and escalating operational expenses. Improving energy efficiency without compromising performance is essential for sustainable HPC operations

Building scalable systems and software architectures for HPC presents numerous challenges and considerations. Among these challenges, load balancing, resilience, and energy efficiency play crucial roles in determining the overall performance and sustainability of HPC architectures.

Here is a comparison of AWS, Azure, and Google Cloud in terms of performance and energy efficiency for HPC workloads:

*1. AWS*

AWS offers specialized Cluster Compute instance types optimized for HPC applications, providing high-performance CPUs, large amounts of RAM, and enhanced networking capabilities. Performance evaluation studies have shown that AWS EC2 HPC instances exhibit equivalent single-node performance compared to traditional supercomputers like NASA's Pleiades cluster. However, when scaling to larger core counts, AWS EC2 HPC instances may lag traditional supercomputers due to network performance differences. AWS provides services like Amazon Elastic Block Store (EBS) and Amazon S3 for scalable storage options, supporting data-intensive HPC applications.

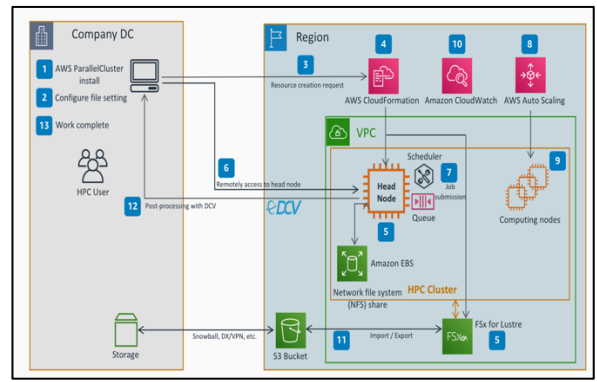

Fig. 3. HPC Reference Architecture on AWS Platform

*2. Azure*

Azure offers HPC-optimized virtual machine series like the H-series and N-series, featuring high-performance CPUs and GPUs for accelerating HPC workloads. Azure VMs with NVIDIA GPUs and InfiniBand interconnects aim to provide high-performance computing capabilities for demanding HPC applications. Azure Blob Storage and Azure Files offer scalable storage options for HPC workloads, ensuring data availability and reliability. Azure Cycle Cloud facilitates the creation and management of HPC and big data clusters in Azure, enhancing the scalability and efficiency of HPC applications.

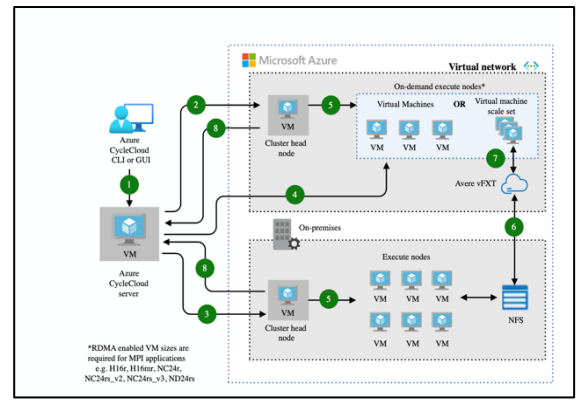

Fig. 4. HPC Reference Architecture on Azure Platform

*3. GCP*

GCP provides high-performance computing instances like the N1 and N2 machine families, offering varying CPU and memory configurations for HPC workloads. GCP supports GPUs from NVIDIA and AMD for accelerating compute-intensive tasks in HPC applications. Google Cloud Storage and Persistent Disk offer scalable storage solutions for HPC workloads, ensuring efficient data management and analysis. GCP supports cluster management tools like Slurm and Kubernetes for job scheduling and resource allocation in HPC environments.

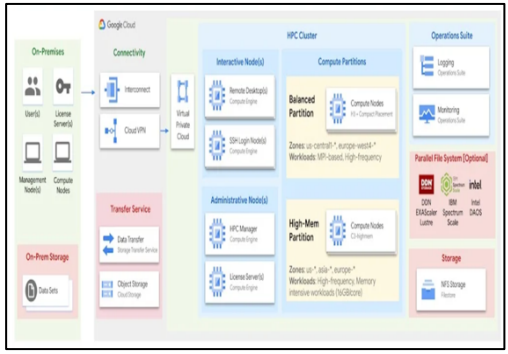

Fig. 5. HPC Reference Architecture on Google Platform

In summary, AWS, Azure, and Google Cloud offer specialized instance types, storage options, and tools tailored for HPC workloads. While each cloud provider has its strengths in terms of performance and scalability, the choice between them may depend on specific workload requirements, network performance considerations, and the need for energy-efficient computing solutions. Further detailed performance benchmarks and energy efficiency studies may provide more insights into the comparative advantages of each cloud platform for HPC applications.

## V. Future Roadmap For Scaling Up HPC Architectures On Cloud Pltforms

High-performance computing (HPC) architectures on cloud platforms such as AWS or Azure are poised for significant advancements, driven by innovations in hardware components such as memory, CPU, GPU, storage, and networking. These advancements will play a pivotal role in scaling up HPC capabilities, enhancing performance, efficiency, and scalability. Here's a future roadmap highlighting key areas of development and their impact on scaling up HPC architectures:

### A. Performance Optimization

Recent advances in CPU architectures, such as higher core counts, improved instruction sets, and increased cache sizes, have significantly boosted computational power. Future advancements, including the adoption of next-generation processors like AMD EPYC Milan or Intel Xeon Scalable processors, promise further performance gains. Additionally, advancements in GPU technology, such as increased CUDA core counts, faster memory interfaces, and specialized tensor cores for AI workloads, will enable HPC architectures to harness the immense parallel processing power of GPUs for scientific simulations and deep learning tasks.

### B. Scalability and Elasticity

Advances in memory technologies, such as high-bandwidth memory (HBM) and non-volatile memory (NVRAM), will enable HPC architectures to scale up memory capacity and bandwidth to accommodate large datasets and complex simulations [3]. Similarly, improvements in storage technologies, including solid-state drives (SSDs), NVM e-based storage, and distributed file systems, will enhance storage performance, scalability, and reliability, enabling HPC architectures to efficiently manage and analyse vast amounts of data [4].

### C. Hybrid and Edge Computing

Advancements in networking technologies, such as high-speed interconnects (e.g., InfiniBand, RDMA over Ethernet) and software-defined networking (SDN) solutions, will enable HPC architectures to achieve low-latency, high-bandwidth communication between compute nodes, facilitating the scalability of large-scale parallel applications. Additionally, innovations in network virtualization and container orchestration platforms, such as Kubernetes and Docker Swarm, will enable seamless integration of HPC workloads with edge computing infrastructure, allowing organizations to leverage distributed computing resources for real-time data processing and analysis.

### D. Data Management and Analytics

Advances in security technologies, including hardware-based encryption, secure enclaves (e.g., Intel SGX, AMD SEV), and trusted execution environments (TEEs), will enhance data protection and confidentiality in multi-tenant cloud environments, ensuring the security and compliance of HPC workloads [7]. Furthermore, developments in compliance automation tools and audit logging frameworks will simplify regulatory compliance for organizations operating in highly regulated industries, such as healthcare, finance, and government.

### E. Quantum and Neuromorphic Computing

Emerging technologies such as quantum computing and neuromorphic computing hold the potential to revolutionize HPC architectures by enabling new computational paradigms for solving complex problems. Cloud providers are investing in quantum computing platforms, such as AWS Bracket and Azure Quantum, to provide developers and researchers with access to quantum hardware and software tools for exploring quantum algorithms and applications. Similarly, advancements in neuromorphic computing, such as IBM's TrueNorth chip, offer new opportunities for energy-efficient, brain-inspired computing architectures that can tackle complex computational tasks with unprecedented efficiency.

In conclusion, advances in hardware components such as memory, CPU, GPU, storage, and networking will play a crucial role in scaling up HPC architectures on cloud platforms. By leveraging these advancements, organizations can achieve higher performance, scalability, and efficiency in their HPC workloads, driving innovation and accelerating scientific discovery across various domains.

## VI. Conculsion

Load balancing, resilience, and energy efficiency are critical considerations in designing scalable HPC architectures, and cloud platforms like AWS and Azure offer a range of services and features to address these challenges effectively. By leveraging dynamic load balancing algorithms, scalable fault tolerance mechanisms, and energy-efficient design practices available on cloud platforms, organizations can build scalable, resilient, and energy-efficient HPC solutions to meet the demands of modern computational workloads while optimizing resource utilization and minimizing operational costs.

REFERENCES

[1] AMD EPYC Milan Processors. (n.d.). Retrieved from https://www.amd.com/en/products/epyc-7003-series-processors.
[2] NVIDIA GPUs for HPC. (n.d.). Retrieved from https://www.nvidia.com/en-us/gpu-accelerated-applications/high-performance-computing/
[3] High-Bandwidth Memory (HBM). (n.d.). Retrieved from https://www.micron.com/products/high-bandwidth-memory
[4] Solid-State Drives (SSDs). (n.d.). Retrieved from https://www.intel.com/content/www/us/en/products/memory-storage/solid-state-drives.html
[5] Mellanox High-Speed Interconnects. (n.d.). Retrieved from https://www.mellanox.com/products/interconnects
[6] Docker Enterprise. (n.d.). Retrieved from https://www.docker.com/products/docker-enterprise
[7] Intel Software Guard Extensions (SGX). (n.d.). Retrieved from https://software.intel.com/content/www/us/en/develop/topics/software-guard-extensions.html
[8] Azure Compliance Offerings. (n.d.). Retrieved from https://azure.microsoft.com/en-us/overview/trusted-cloud/compliance/
[9] AWS Quantum Computing. (n.d.). Retrieved from https://aws.amazon.com/braket/
[10] IBM TrueNorth Chip. (n.d.). Retrieved from https://www.research.ibm.com/artificial-intelligence/truenorth/
[11] J. E. Simons, "Virtualization for HPC," 2012 IEEE International Symposium on Workload Characterization (IISWC), La Jolla, CA, 2012, pp. 116-117.
[12] D. B. Muralitharan, S. A. B. Reebha and D. Saravanan, "Optimization of performance and scheduling of HPC applications in cloud using cloudsim and scheduling approach," 2017 International Conference on IoT and Application (ICIOT), Nagapattinam, 2017, pp. 1-6.
[13] D. Beserra, M. K. Pinheiro, C. Souveyet, L. A. Steffenel and E. D. Moreno, "Performance Evaluation of OS-Level Virtualization Solutions for HPC Purposes on SoC-Based Systems," 2017 IEEE 31st International Conference on Advanced Information Networking and Applications (AINA), Taipei, 2017, pp. 363-370.
[14] D. Beserra, E. D. Moreno, P. T. Endo and J. Barreto, "Performance evaluation of a lightweight virtualization solution for HPC I/O scenarios," 2016 IEEE International Conference on Systems, Man, and Cybernetics (SMC), Budapest, 2016, pp. 004681-004686.
[15] S. Iserte, J. Prades, C. Reaño and F. Silla, "Increasing the Performance of Data Centers by Combining Remote GPU Virtualization with Slurm," 2016 16th IEEE/ACM International Symposium on Cluster, Cloud and Grid Computing (CCGrid), Cartagena, 2016, pp. 98-101.
[16] . Zhang, X. Lu and D. K. Panda, "Performance Characterization of Hypervisor-and Container-Based Virtualization for HPC on SR-IOV Enabled InfiniBand Clusters," 2016 IEEE International Parallel and Distributed Processing Symposium Workshops (IPDPSW), Chicago, IL, 2016, pp. 1777-1784.